\documentclass[10pt,letterpaper,twocolumn]{article} 
\usepackage{ol2}
\usepackage[draft]{hyperref}
\usepackage{amsmath}
\usepackage{subfigure}
\usepackage{graphicx}
\usepackage{cite}
\begin{document}

\twocolumn [

\title{Extraordinary Optical Transmission: Coupling of the Wood-Rayleigh anomaly and the Fabry-Perot resonance}


\author{A. T. M. Anishur Rahman$^{*}$, Peter Majewski and Krasimir Vasilev}
\address{School of AME, University of South Australia\\ Mawson Lakes, SA 5095, Australia}
\email{$^*$ Corresponding Author: rahaa001@mymail.unisa.edu.au} 

\begin{abstract}In this letter, we demonstrate for the first time that by combining the effects of the Wood-Rayleigh Anomaly (WRA) and the Fabry-Perot (FP) resonance, transmission efficiencies of $1$D metallo-dielectric gratings on substrates can be significantly improved compared to when these two phenomena work separately. Results of combining the WRA and the FP resonance can be utilized to eliminate the necessity of using the index matching technique and the core-shell structure for enhancing the performance of extraordinary optical transmission devices. Further, the outcomes of combining the WRA and the FP resonance can elucidate some of the unexplained results in the literature.\end{abstract}

] 

\begin{figure}
\centering
\subfigure{\includegraphics[width=6.0cm]{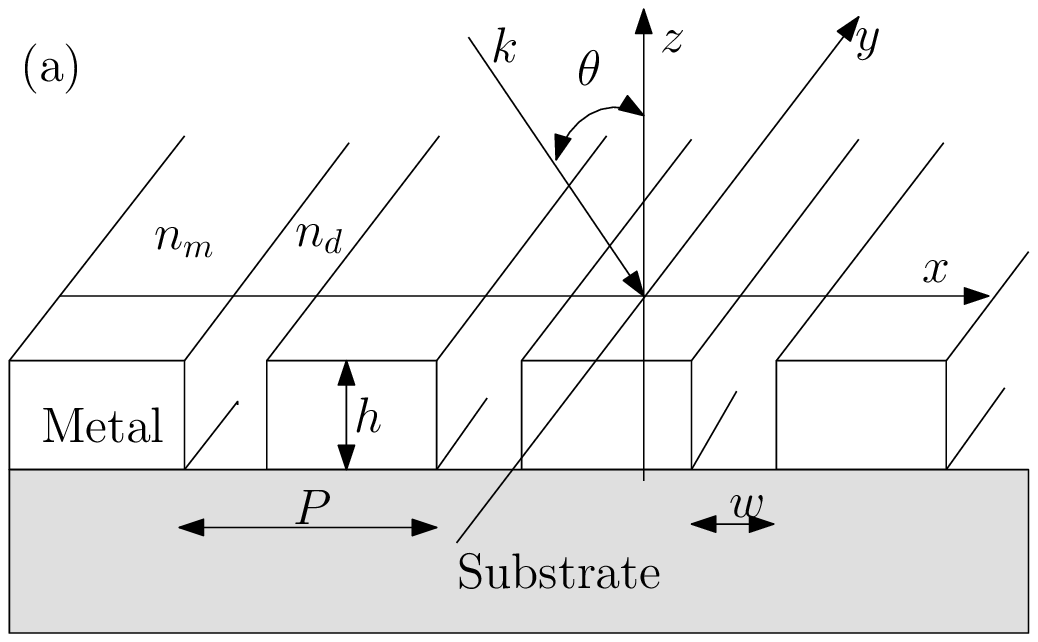}}\\
\subfigure{\includegraphics[width=7.0cm]{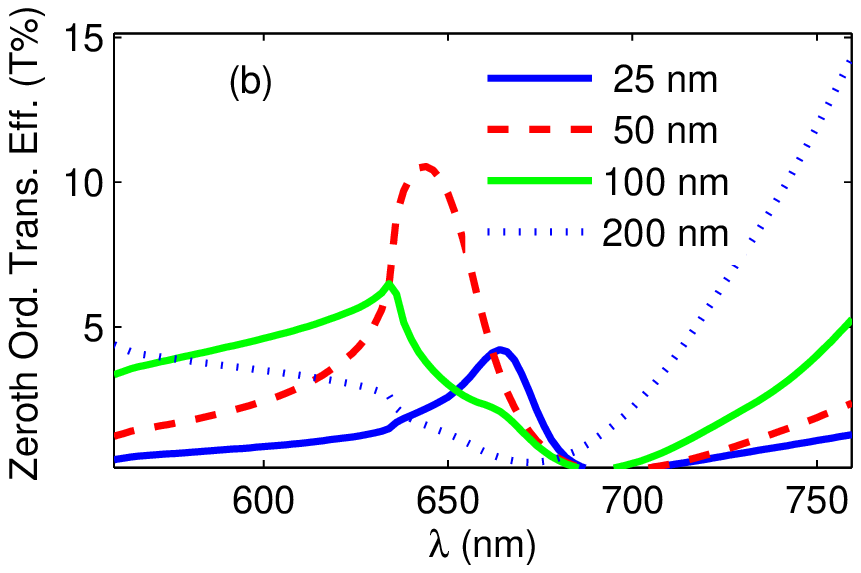}}
\centering
\caption{(a) $1$D grating structure. Assuming a gold \cite{Palik85} grating on a glass substrate ($n_s=1.52$), (b) effects of the slit width on the desired peak at $\lambda=635$ nm for $P=418$ nm, $h=209$ nm, $m=0$ and $\theta=0^o$.}
\label{fig1}
\end{figure}

After the pioneering discovery of Extraordinary Optical Transmission (EOT) through $2$D metallic hole arrays \cite{Ebbesen1998}, Porto et al. \cite{Porto1999} demonstrated that $1$D metallo-dielectric gratings with sub-wavelength slits ($w < \lambda/2$, see Fig. \ref{fig1}) can also show similar effects. Subsequently $1$D metallo-dielectric gratings have been proposed for many novel applications including Near-field Scanning Optical Microscopy (NSOM), spatial light modulators and flat panel displays \cite {Astilean2000,Kim1999}. It is believed that when the wavelength ($\lambda$) of EOT is equivalent to the Rayleigh wavelength, the Wood-Rayleigh Anomaly (WRA) is responsible for EOT \cite{Garcia2007,Steele2003} while for deep enough gratings and $\lambda >  P$, where $P$ is the grating period, the Fabry-Perot (FP) resonance mechanism is considered to be accountable for the enhanced transmission through $1$D metallo-dielectric grating structures \cite{Porto1999, Cao2002,Rahman2012}.

In this letter, we show that by combining the effects of the WRA and the FP resonance (traditionally considered to be responsible for EOT at different regimes of wavelengths) the transmission efficiency of a $1$D grating on a substrate can almost be doubled compared to the cases when the WRA and the FP resonance work separately. The possibility of coupling the WRA and the FP resonance have been predicted before \cite{ Marquier2005,DAguanno2011} but it has never been used to enhance the performance of an EOT device. Enhanced performance of an EOT device due to the combined effects of the WRA and the FP resonance can replace the index matching technique \cite{Krishnan2001} (normally achieved by mixing liquids of different refractive indexes and hence is not feasible for many real applications) for boosting the efficiency of a $1$D EOT device on a substrate. Our approach can also substitute the core-shell structure \cite{Collin2010} which is normally used to by-pass the adverse effects of a substrate on  EOT (not feasible for visible band operation due to the large metallic wall thickness to get a reasonable mechanical strength and the associated ohmic loss). Further, outcomes of combing the effects of the WRA and the FP resonance can interpret some of the unexplained results in the original article of Porto et al. \cite{Porto1999} on EOT. In particular, Porto and co-workers reported that as the grating thickness increases from $0.60$ $\mu$m to $3$ $\mu$m, the transmission peak related to the Surface Plasmon Polariton (SPP) (equivalently due to the WRA) close to $\lambda=3.5$ $\mu$m ($P=3.50$ $\mu$m, $w=0.50$ $\mu$m, see Fig. $1$, Ref. $2$) shifts towards longer wavelengths. Although, the SPP resonance does not depend on the grating thickness \cite{Ebbesen1998, Ghaemi1998}. In the same article, when the photonic band diagram (Fig. $2$ (b), Ref. $2$) for $P=3.50$ $\mu$m, $w=0.50$ $\mu$m and $h=3.00$ $\mu$m was plotted, a flat band at $\approx0.35$ eV appeared in addition to the SPP bands; even though the relevant transmission peak at $\lambda\approx 4.00$ $\mu$m was assigned to the surface plasmon polariton by Porto et al. \cite{Porto1999}. The source of this flat band is unknown. However, according to us, all of these phenomena are due to the interaction of the SPP/WRA and the FP resonance (see below for details).

Let us assume that an electromagnetic wave of free space wavelength ($\lambda$) is incident upon the grating structure of Fig.
\ref{fig1} (a) at an incidence angle ($\theta$). Further $w < \lambda/2$ and the EOT is supposed to be channeled to the $m^{th}$ ($m=0,1,2,3,..$) transmitted diffraction order are also assumed. To attain enhanced transmission at a desired $\lambda$ exploiting the Wood-Rayleigh anomaly, Eq. ($1$) needs to be fulfilled \cite{Hessel1965}, where $n_1$ and $n_s$ are the refractive indexes of the incidence medium and the substrate respectively. This condition ensures that when the $(m+1)^{th}$ diffraction order dies off at the desired $\lambda$, associated energy of this order is distributed among the propagating orders and a transmission peak appears \cite{Hessel1965}. Now, to warrant a coupling between the $m^{th}$ propagating order in the transmission media and a waveguide mode (here the fundamental mode since $w < \lambda/2$), the condition of Eq. ($2$) needs to be satisfied \cite{Clausnitzer2005}. In Eq. ($2$), all parameters are known except $n_{Eff}$ which is the effective index of a slit corresponding to the fundamental waveguide mode. $n_{Eff}$ can be expressed as - $n_{Eff}^2=1+i\lambda/(\pi w n_m)$ where $n_m$ is the refractive index of the ridge material \cite{Rahman2011A}. Since $Re(n_{Eff}^2) >> Im(n_{Eff}^2)$ \cite{Rahman2011A}, $Re(n_{Eff})^2$ can be approximated by $Re(n_{Eff}^2)$. Substituting $Re(n_{Eff})^2$ in Eq. ($2$), one can obtain $w$ from Eq. ($3$), where $\eta$ and $\kappa$ are the real and the imaginary components of $n_m$. $n_m^*$ is the conjugate of $n_m$. Lastly, to ensure that the FP resonance condition is also met at the wavelength of the WRA, Eq. ($4$) needs to be satisfied \cite{Astilean2000}. Eq. ($4$) confirms that the light subjected to multiple reflection inside a slit cavity interferes constructively at the $z=-h$ interface. Refractive index of the slit is assumed to be $1$. Optimum grating thickness can be found from Eq. ($4$).

\begin{eqnarray}
P=\frac{(m+1)\lambda}{n_s-n_1\sin{\theta}}\\
k_0^2Re(n_{Eff})^2-k_0^2n_s^2+(k_0n_1\sin{\theta}+2\pi m/P)^2=0\\
w=\frac{\lambda \kappa}{\pi[n_s^2-1-(n_1\sin{\theta}+m\frac{\lambda}{P})^2]n_mn_m^*}\\
h=\frac{j\lambda}{2Re(n_{Eff})}
\end{eqnarray}

\begin{figure}
\centering
\subfigure{
\includegraphics[width=7.0cm]{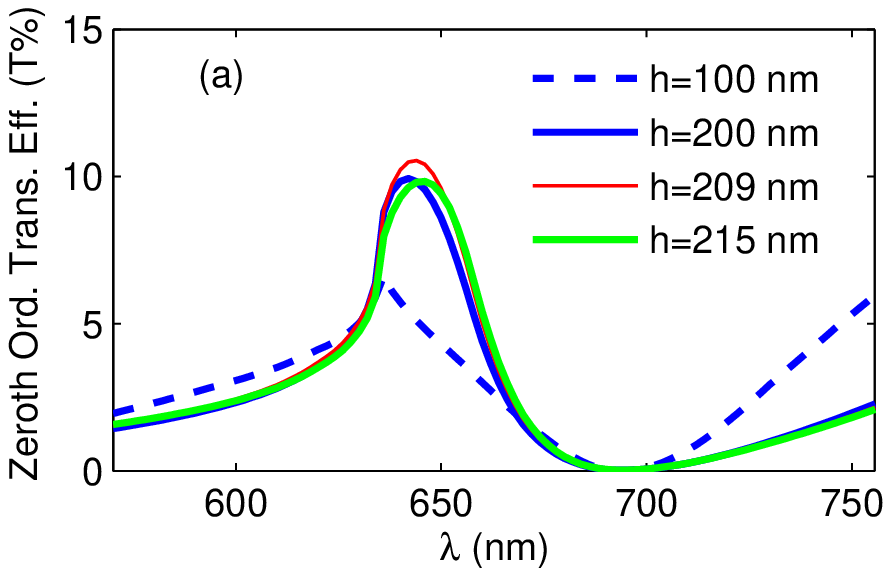}}
\subfigure{
\includegraphics[width=7.0cm]{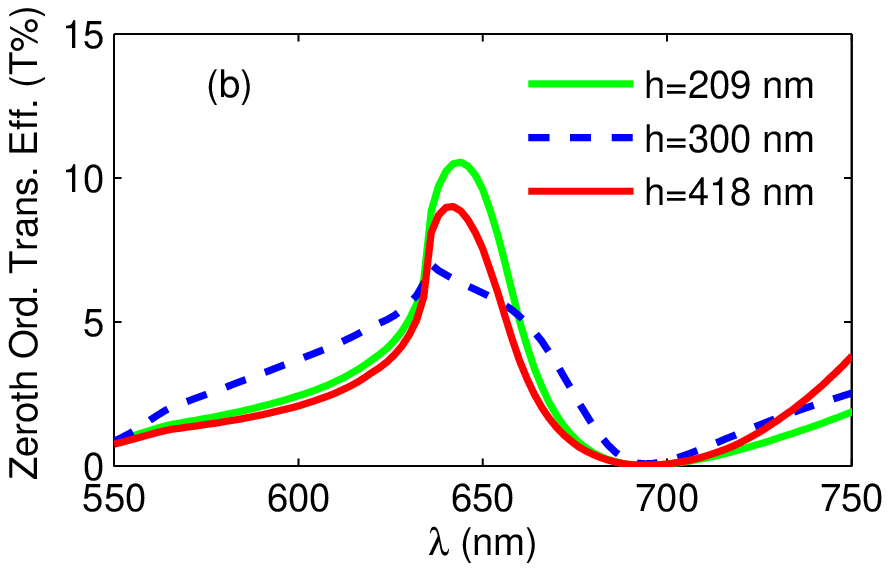}}
\centering \caption{Assuming a gold grating (a) and (b) show the impact of the grating thickness on the transmission peak while the substrate ($n_s=1.52$, glass), the grating period ($P=418$ nm) and the slit width ($w=50$ nm) are held  constant.}
\label{fig4}
\end{figure}

Assuming a red color filter is desired at $\lambda=635$ nm \cite{Kim1999}, for $\theta=0^o$ and $m=0$, one gets $P=418$ nm, $w=50$ nm and $h=209$ nm (see equations ($1$)- ($4$)). Fig. \ref{fig1} (b) shows transmission spectrum obtained using the Rigorous Coupled Wave Analysis (RCWA) \cite{Moharam1995} corresponding to the previous parameters. A well-defined transmission peak of small line-width appears at $644$ nm. To confirm the role of the FP resonance, transmission spectra corresponding to slit widths $25$ nm, $100$ nm and $200$ nm have also been included in Fig. \ref{fig1} (b). Since $n_{Eff}$ varies inversely with $w$, these values of slit widths satisfy the FP resonance conditions (Eq. ($4$)) far away from the WRA condition. Consequently, transmission peaks only due to the Wood-Rayleigh anomaly close to the desired wavelength are expected to be present. This is precisely the case in Fig. \ref{fig1} (b). Also, in all cases peak transmission intensities are less than or equal to the half of the optimum case ($w=50$ nm). This result shows the potential of using the combined effects of the FP and WRA mechanisms in replacing the index matching and core-shell techniques of boosting performances of EOT devices on substrates. To demonstrate further that the FP resonance can be utilized in combination with the WRA to boost transmission intensity at a desired wavelength, Fig. \ref{fig4} shows the effects of changing the grating thickness while keeping all other parameters unchanged. Since $w$ and $P$ are constants ($n_{Eff}$ is also constant), changing $h$ means that the Fabry-Perot resonance occurs at a different wavelength than the Rayleigh wavelength. Consequently, the combined effects of the two mechanisms should be lost and the relevant peak intensity is supposed to be diminished. This is true in Fig. \ref{fig4} (a) when $h$ differs substantially (i.e. $h=100$ nm and $h=300$ nm) from the optimum grating thickness ($h=209$ nm) or the FP resonances occur far away from the Wood-Rayleigh anomaly. However, when $h$ differs by a multiple of the optimum grating thickness i.e. $h=418$ nm, peak intensity and profile almost go back to that of the optimum case ($a=50$ nm, $h=209$ nm and $P=418$ nm). This is because for $w=50$ nm, $h=418$ nm and $P=418$ nm, the Fabry-Perot condition is again satisfied at $\lambda=635$ nm (see Eq. ($4$) with $j=2$) resulting in an enhanced transmission in addition to that resulted from the WRA. Slightly lower peak intensity for $h=418$ nm compared to that for $h=209$ nm in Fig. \ref{fig4}(b) can be explained as the ohmic loss incurred by a waveguide mode when it travels a longer optical path through a real metallic guide. Further, in Fig. \ref{fig4}(a) when $h$ differs by a small amount from the optimum value such as $h=200$ nm and $h=215$ nm, relative changes in the peak intensities and profiles are very little. This should provide the necessary tolerance in the actual device fabrication for extraordinary optical transmission.

\begin{figure}
\centering
\includegraphics[width=7.0cm]{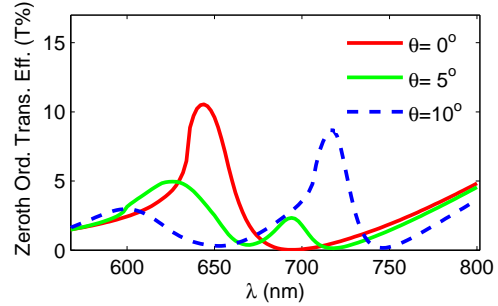}
\centering
\caption{Impact of the incidence angle on the combined transmission peak due to the WRA and the FP for $w=50$ nm, $h=209$ nm and $P=418$ nm.}
\label{fig5}
\end{figure}

Finally, to consolidate our claim that the WRA and the FP resonance can be combined at a wavelength to enhance performances of EOT devices on substrates, Fig. \ref{fig5} shows transmission spectra for $\theta=0^o$, $\theta=5^o$ and $\theta=10^o$. The dependence of the FP resonance on $\theta$ is minimal \cite{Porto1999} while the WRA depends on it substantially (see Eq. ($1$)). Transmission peaks due to $m=-1$ disappearing beams of the transmission side for $\theta=5^o$ and $\theta=10^o$ appear at $694$ nm and $718$ nm respectively. In the high energy side of Fig. \ref{fig5}, the scenario is much more complicated due to the possible interaction of the Wood-Rayleigh anomalies of $m=-1$ of the incident side, $m=+1$ of the transmission side and the FP resonance of the slit cavity. Nevertheless, transmission peaks with much broader line-widths (characteristics of the FP resonances) appear at $624$ nm and $600$ nm for $\theta=5^o$ and $\theta=10^o$, respectively. Upon close inspection, shoulders at $\lambda=600$ nm for $\theta=5^0$ and at $\lambda=564$ nm for $\theta=10^o$ due to $m=+1$ WRAs of the transmission side can be found. It is noticeable that in all cases where the FP resonance and the WRA are located at different wavelengths, peak transmission intensities are lower than when these two mechanisms are collocated at a wavelength. This again confirms the possibility of using the combined effects of the Wood-Rayleigh anomaly and the Fabry-Perot resonance in replacing the inconvenient index matching technique and the core-shell structure for enhancing the performance of EOT devices. Further, from the result of this section, the origin of the transmission peak at $\lambda\approx 4.0$ $\mu$m for $h=3.00$ $\mu$m, $w=0.50$ $\mu$m and $P=3.50$ $\mu$m in Ref. $2$ can be considered to be the FP resonance plus the WRA or SPP; instead of the SPP alone as assigned by Porto et al. \cite{Porto1999}. Alternatively, the source of the flat band in Fig. $2$ (b) in \cite{Porto1999} is the Fabry-Perot resonance. Before concluding it is important to note that the extraordinary optical transmission can be coupled to higher diffraction orders in addition to the zeroth order by choosing $m>0$ in equations ($1$)-($4$). This can be useful in achieving wide viewing angle of flat panel displays.

In conclusion, we have demonstrated that by combining the Wood-Rayleigh anomaly and the Fabry-Perot resonance, the transmission efficiency of a $1$D metallo-dielectric grating can be significantly improved compared to when the WRA and the FP mechanisms work alone. It has also been argued that the combined effects of the WRA and FP resonance mechanisms can be used to replace the existing index matching and core-shell techniques for boosting transmission efficiencies of EOT devices.

\pagebreak

\title{Extraordinary Optical Transmission: Coupling of the Wood-Rayleigh anomaly and the Fabry-Perot resonance}

\end{document}